\documentclass[useAMS,usenatbib,fleqn]{mn2e}
\usepackage{graphics,graphicx,times}
\usepackage{amsmath}
\usepackage{amssymb}
\usepackage{aas_macros}
\usepackage[usenames,dvipsnames]{xcolor}
\newcommand{\kms}{${\rm km\ s}^{-1}$\ }
\def\mgtwo{Mg\,{\sc ii}\ }

\title{Signature of outflows in strong \mgtwo absorbers in quasar sightlines} 

\author[Sharma et al.]{Mahavir Sharma$^{1,3}$, Biman B. Nath$^{1,4}$ and Hum Chand$^{2,5}$\\
  $^{1}$ Raman Research Institute, Sadashivanagar, Bangalore 560080, India\\
  $^{2}$ Aryabhatta Research Institute of Observational Sciences, Manora Peak, Nainital 263129, India\\
  $^{3}$ mahavir@rri.res.in\,,
  $^{4}$ biman@rri.res.in\,,   
  $^{5}$ hum@aries.res.in
} 
\begin{document}

\date{Submitted --------- ; in original form ---------}

\pagerange{\pageref{firstpage}--\pageref{lastpage}} 

\maketitle

\label{firstpage}
\begin{abstract}
We report a correlation between velocity offset ($\beta=v/c$) of strong \mgtwo absorption systems and the bolometric luminosity ($L_{\rm bol}$) of quasars in SDSS-DR7. We find that,  $\beta$ shows a power law increase with $L_{\rm bol}$, with a slope $\sim 1/4$. We find that such a relation of $\beta$ with $L_{\rm bol}$ is expected for outflows driven by scattering of black hole radiation by dust grains, and which are launched from the innermost dust survival radius. Our results indicate that a significant fraction of the strong \mgtwo absorbers, in the range of $\beta=0\hbox{--}0.4$ may be associated with the quasars themselves.
\end{abstract}
\begin{keywords}
galaxies : active --  quasars : general -- quasars : absorption lines  
\end{keywords}
\section{Introduction}
The study of \mgtwo absorption line systems in the spectra of quasars (QSOs) has
been  useful in detecting distant normal field galaxies situated close to the lines of sight of QSOs
\citep{1991A&A...243..344B,1994ApJ...437L..75S}. Conventionally,
all such absorbers with
velocity $<5000$ \kms relative to the background QSO are believed to
be associated to the QSO itself (`associated systems') while those at
larger velocity offset are believed to be entirely independent of
background QSO. 
This general belief was questioned recently by
the puzzling results on the abundance of strong \mgtwo absorber having
equivalent width ($W_r$) more than 1.0 \AA\ : (i) by
\citet{Prochter2006ApJ...648L..93P} where they found $2\hbox{--}4$ time excess of strong
\mgtwo absorber towards the $\gamma$ ray burst (GRB) sources relative to QSO sight
lines (see also \citealt{2007ApJ...669..741S,Vergani2009A&A...503..771V,Tejos2009ApJ...706.1309T}), and (ii) by 
\citet[]{Bergeron2011A&A...525A..51B},
where they found similar excess by a factor of about 2 (3 $\sigma$
confidence) towards 45 blazar sight lines. 

These counter-intuitive
results, have inspired many alternative explanations, such as dust
extinction towards QSO sight lines which can lower the apparent
incidence rate of absorbers, or gravitational lensing which can
increase it toward GRBs/blazars, but none have been found to explain
the above discrepancies \citep{Porciani2007ApJ...659..218P,Menard2008MNRAS.385.1053M,Lawther2012A&A...546A..67L}.
 However the blazars, as a
class, are believed to have relativistic jet pointed close to our line of sight.
\cite{Bergeron2011A&A...525A..51B} speculated that such powerful jets in the blazars are
capable of sweeping sufficiently large column densities of gas (up to
$10^{18}\hbox{--}10^{20} {\rm cm}^{-2}$ ) and accelerating such clouds to velocities of
order $\sim 0.1c$, thereby possibly accounting for the excess of \mgtwo
absorption systems towards blazars in comparison with QSOs. However, such an
excess in number of \mgtwo absorbers per unit redshift ($dN/dz$) was not confirmed in the analysis of flat-spectrum radio
quasars (FSRQs) by  \citet{Chand2012ApJ...754...38C}, though FSRQs also 
possess powerful jets similar to blazars, which they reconciled with
the above hypothesis of relativistically ejected absorbing clouds, by suggesting that perhaps due to larger angle from the line
of sight (unlike blazars with smaller angle), these accelerated clouds
might not intersect the line-of-sight in the case of FSRQs. 
Using a  larger sample size of 95 GRB (including 12 GRB from
\citealt{Prochter2006ApJ...648L..93P}), \citet{Cucchiara2012arXiv1211.6528C} did not confirm the
original enhancement found in the case of GRB by \citet{Prochter2006ApJ...648L..93P}, though a signature of mild excess of about 1.5
times was noticed for strong \mgtwo absorption systems, albeit with
only a low confidence level of 90\%.

The firm conclusion for jet based above excess  still await the
realistic numerical modelling of jet-ambient gas interaction especially
for the excess seen towards blazars (about a factor of 2) and CDQs (about
10\%) \citep{Joshi_etal} vis-a-vis normal QSOs. 
However an alternative scenario, which could be more plausible, is the dust or radiation
driven outflows \citep[e.g.][]{1995ApJ...451..510S}.
For instance, if there is some contribution to $dN/dz$ of
strong \mgtwo absorber from these outflows, then one will expect that
AGN luminosity should have statistical correlation with the velocity
offset of the strong \mgtwo absorber relative to the background AGN,
which is usually defined by,
\begin{equation}
\beta = \frac{(1+z_{\rm qso})^2-(1+z_{\rm abs})^2}{(1+z_{\rm qso})^2+(1+z_{\rm abs})^2}
\end{equation}
where $\beta=v/c$, $z_{\rm qso}$ is the emission redshift of the QSO and $z_{\rm abs}$ is the absorption redshift of the   \mgtwo system.

In this letter we report a correlation between the $\beta$ of strong \mgtwo absorbers and the bolometric luminosity ($L_{\rm bol}$) of QSOs,  using the strong \mgtwo absorber catalogue by \citet{Lawther2012A&A...546A..67L}. We also propose an explanation for this correlation which draws upon radiation driven outflow models. In \S 2 we describe the sample  of strong \mgtwo absorbers and our selection criteria. In \S 3 we present our results and a theoretical model of radiation driven outflows. In \S 4 we study the fractional number counts of absorbers, and discuss our results in \S 5.
\section{Description of the sample}
We consider a sample of 10367 strong \mgtwo absorbers with equivalent
width $W_r(2796) > 1$\AA\ belonging to 9144 QSOs, from the recent
compilation by \citet{Lawther2012A&A...546A..67L} based on 105783 QSOs 
of SDSS DR7 \citep{Abazajian2009ApJS..182..543A,Schneider2010yCat.7260....0S}. However, the
range of $\beta$ varies with 
$z_{\rm qso}$, and the observed wavelength range of the spectrum. 
Therefore, in order to make the sample unbiased,  
firstly, we have considered a SDSS
spectral range from 4000-9000 \AA\, which is a little narrower (by about 100 \AA) than the  actual one.
We then applied the following four selection filters. 
\begin{enumerate}
\item We removed 773 broad absorption line (BAL) QSOs 
from our above sample to avoid any contamination in our analysis by BAL features which 
has resulted in the removal of  corresponding 931 strong \mgtwo absorbers.

\item For all the quasars having $z_{\rm qso} > 2.21024$, the \mgtwo 
emission line will fall above 9000 \AA, which is our conservative upper limit on wavelength of 
SDSS spectrum. As a result, SDSS spectra for such 
sources will not allow any detection of strong \mgtwo doublet falling 
in the redshift range between 2.21024 up to $z_{\rm qso}$. Therefore to avoid 
this observational bias, we excluded all sources having $z_{\rm qso} \ge 2.21024$ 
from our sample, which resulted in the removal of 43 QSOs having 52 strong
Mg II absorbers.

\item  Another filter was applied to avoid the observational bias 
which might result from the lower wavelength limit, viz 4000 \AA, in 
the SDSS spectra. In our analysis we aim to see any correlation of luminosity 
with velocity offset up to about 0.4c. However for 4000 \AA\ considered as 
the conservative starting wavelength of our spectra, $z_{\rm qso} = 1.185$ will be the minimum redshift, which allows us to detect \mgtwo 
absorber (if any) at least up to a velocity offset of $0.4 c$. Therefore, we have removed 1461 
sightlines with $z_{\rm qso} < 1.185$ having 1544 strong \mgtwo absorbers in their spectra.

\item  After applying the above mentioned redshift cuts, we are left with 
the systems with $2.21024>z_{\rm qso}\geq 1.185$. In these intermediate 
redshift systems, the $\beta$ value can be larger than $0.4$, which in principle may give rise to a bias of higher $\beta$ with increasing $z_{\rm qso}$. 
Hence we  also remove all the absorbers with $\beta>0.4$ from the remaining sample which amounts to exclusion of 1523 absorbers along 1439 sightlines. 
One should note that $\beta =0.4$ is chosen because if we keep $\beta$ value less or greater than this, then the sample is significantly reduced. Another motivation as will be clear in the coming sections, is that $\beta \sim 0.4$ is an upper limit for the radiation (dust) driven outflows.
\end{enumerate}
Finally, we are left with
6317 strong \mgtwo systems along 5682 QSOs in the selected redshift range. 
Bolometric luminosities  for the QSOs in SDSS DR7 are calculated in a recent 
study by \citet{shen2011ApJS..194...45S}. We cross matched the QSOs in our sample, with the catalogue described in Shen's paper to obtain the  bolometric luminosity. We then removed two more absorbers whose QSO luminosities were $< 10^{45}$ erg s$^{-1}$. Our final bias free sample consists of 6315 strong \mgtwo systems with luminosity
range $10^{45.5} < L_{\rm bol} \le 10^{47.8}$ erg s$^{-1}$, with redshift range $1.185 \le 
z_{\rm qso} < 2.21024$, and with the velocity offset range of 
$0 < \beta c \le 0.4 c$.
 In Figure 1, the blue dashed
 line represents the distribution of strong Mg II absorbers in
 SDSS-DR7, compiled by \citet{Lawther2012A&A...546A..67L}, and the black solid line is the
 final sample selected for this study. 
\begin{figure}
\includegraphics[scale=0.4]{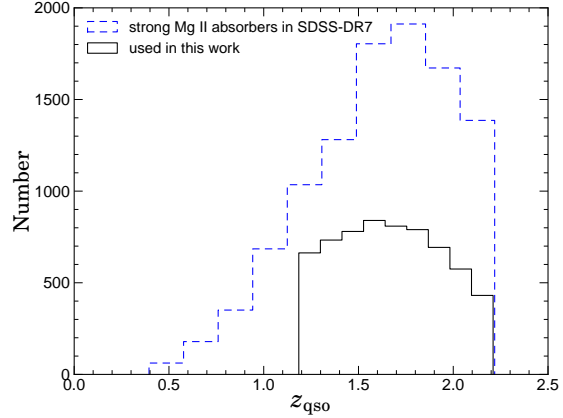}
\caption{Histograms with $z_{\rm qso}$ of the samples of strong \mgtwo absorbers in SDSS-DR7. The blue dashed line is for 10367 strong absorbers compiled by \citet{Lawther2012A&A...546A..67L}. The black solid line represents the sample used in this work.}
\end{figure}
\section{Correlation between $\beta$ and $L_{\lowercase{\rm bol}}$ : signature of radiation driven outflow}
In order to test the dependence of $\beta$ on luminosity, we  divide the sample in bins of bolometric luminosity. Most of the absorbers (5651 out of 6315) belong to QSO sightlines having a luminosity range $10^{46}\hbox{--}10^{47}$ erg s$^{-1}$. We divide these 5651 systems into four bins of bolometric luminosity. We also have two more bins, one for $L_{\rm bol}<10^{46}$ erg s$^{-1}$, and another with $L_{\rm bol}>10^{47}$ erg s$^{-1}$, the first having 27 systems and the second with 637 systems.

Consider the case of the absorbers being distributed uniformly in the allowed range of $z_{\rm abs}$ (which in turn is determined from
the allowed range of $\beta$), then the median value of $\beta$ should be independent of $z_{\rm qso}$ (see Appendix A for a proof). Hence, irrespective of the distribution of $z_{\rm qso}$ in a luminosity bin, the median value of beta should be same in all luminosity bins.
To test this hypothesis, we estimate the median, the lower 25 percentile and the upper 25 percentile of data in each of the above mentioned six luminosity bins. We plot the median with  circles, and  the upper and lower percentiles as the end points of vertical dotted bars in Figure 2.
\begin{figure}
\includegraphics[scale=0.4]{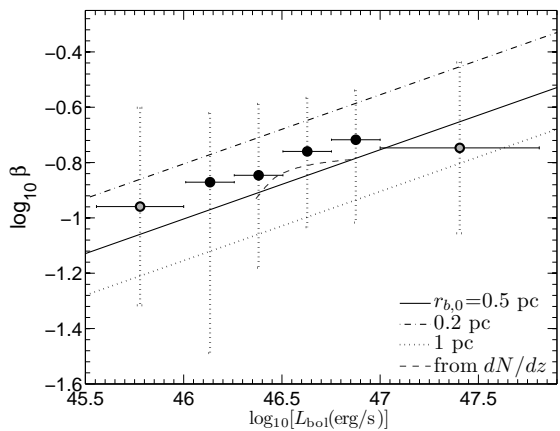}
\caption{Correlation between the bolometric luminosity of the QSO and $\beta$ of \mgtwo absorber. The  circles represent  the median of data in a particular luminosity bin. The upper and lower extreme of the dotted vertical lines give the location of upper and lower 25 percentile, respectively. Sizes of the luminosity bins are shown by the horizontal bars. The solid, dotted and dash-dotted line represent the theoretical model discussed in \S 3.1.  }
\end{figure} 

Interestingly, we find that the median is not constant. The data shows a correlation of $\beta$ with the $L_{\rm bol}$. The 5651 absorbers systems with $10^{46}\leq L_{\rm bol}\leq 10^{47}$ erg s$^{-1}$, which form the mainstay of the sample show a power law increase of $\beta$ with $L_{\rm bol}$, with a slope of $\sim1/4$. 
Increase of median value of $\beta$ with  the bolometric luminosity, proves that the distribution in each bin is not uniform random.
This fact is also hinted in the evolution of $dN/dz$ with $z_{\rm abs}$ \citep{Zhu2012arXiv1211.6215Z}.
Using the evolution in $dN/dz$ for our sample, we evaluated the expected relation between the median value of $\beta$ and $z_{\rm qso}$ from equation (\ref{app_med}). We then converted it to the corresponding relation between $\beta$ and luminosity by using the best fit relation between $z_{\rm qso}$ and luminosity, a characteristic of  magnitude limited survey such as  SDSS. We have shown this relation using a dashed curve in Figure 2. 
Although the dashed curve  does show some evolution, but it is clear that it  
cannot fully explain the observed $\beta\hbox{--}L_{\rm bol}$ correlation.

Which physical processes can give rise to non-uniformity of absorber distribution? The evolution in $dN/dz$ has been attributed to the evolution in global star formation rate \citep{Zhu2012arXiv1211.6215Z}, although without any concrete evidence. 
Also, observations of intervening galaxies 
show a small covering fraction ($\le 0.3$) for strong \mgtwo absorbers ($W \ge 1$ \AA)
\citep{2012arXiv1211.1380N,2010ApJ...714.1521C}.
Here we explore an alternate  based on outflows associated with QSOs, which can give rise to non-uniformity of incidence of absorbers. As we explore in next section, the relation $\beta \propto L_{\rm bol}^{1/4}$, is a natural consequence of QSO radiation driven outflows.
\subsection{Absorbers as radiation driven outflows}
Radiation driven outflows have been invoked repeatedly in literature to explain the co-evolution of black hole and bulge, to explain the accretion disc winds \citep[e.g.][]{2000ApJ...543..686P} and galactic winds \citep[e.g.][]{2005ApJ...618..569M,2011ApJ...736L..27S}. We consider here the radiation driven outflows, where the photons scatter the dust grains and impart their momentum to dust. The dust in turn is collisionally coupled to the gas, and the momentum is uniformly distributed to the dust and gas mixture. In this scenario, the motion of  dust and gas mixture surrounding the QSO is governed by the following equation,
\begin{equation}
v{dv \over dr} = {\kappa L_{\rm uv} \over 4\pi r^2 c} - {G M_\bullet \over r^2} - {d\Phi \over dr}
\label{eq_diff}
\end{equation}
where $M_\bullet$ is the mass of the black bole and $\Phi$ is the dark matter halo potential. $L_{\rm uv}$ is the integrated UV luminosity and for QSOs where the main emission is in high frequency  bands,  luminosity over UV and EUV bands is roughly half of the bolometric luminosity ($L_{\rm  uv}\sim0.5 L_{\rm bol}$)\citep{2004ASSL..308..187R}. $\kappa$ is the frequency averaged opacity for the scattering and absorption of UV photons by dust grains. For wavelength of photon $<0.3\ \mu$m, the $\kappa$ for a dust and gas mixture ranges  from 200 to as large as 1000 cm$^{2}$g$^{-1}$ \citep{2001ApJ...554..778L}. We take a value $\kappa=500$ cm$^{2}$g$^{-1}$, which roughly serves as an average effective value of opacity.
We can integrate equation (\ref{eq_diff}) to obtain the following expression for velocity
\begin{equation}
v^2 = {\kappa L_{\rm bol} \over 4 \pi c}\left({1 \over r_b} - {1\over r}\right) - 2(\Phi(r)-\Phi(r_b))%
\end{equation}
where $r_b$ is the launching radius of the outflow. In the case of  radiation pressure on dust grains, the opacity is generally quite high and hence the  radiation force is many times larger than the gravity, therefore the gravitational force can be neglected.  At a large  distance the velocity attains the following terminal value
\begin{equation}
v_{\infty} \simeq \left({\kappa L_{\rm bol} \over 4\pi c\ r_b}\right)^{1/2}
\label{wind_easy}
\end{equation}
The base radius ($r_b$) for launching these outflows is an important factor and it should be the minimum distance at which the dust grains can survive. Studies on dust survival yield following relation between the  sublimation radius of the dust grains  and the luminosity of the AGN \citep{2012MNRAS.420..526M},
\begin{equation}
r_b = R_{\rm sub} \sim r_{b,0}\ \left(\frac{L_{\rm bol}}{10^{46}\ {\rm erg\ s^{-1}}}\right)^{0.5} \,.
\label{eq_mor} 
\end{equation}
The value of $r_{b,0}$ is $0.5$ pc for graphite grains and $1.3$ pc for the silicate grains. Substituting equation (\ref{eq_mor}) into (\ref{wind_easy}), we obtain the following expression for wind terminal speed, 
\begin{equation}
v_\infty \sim 0.1 c\ \left(\frac{\kappa}{500\ {\rm cm^2\ g^{-1}}}\right)^{1/2}  \left(\frac{L_{\rm bol}}{10^{46}\ {\rm erg\ s^{-1}}}\right)^{1/4}\left(\frac{r_{b,0}}{{0.5\ {\rm pc}}}\right)^{-1/2}
\end{equation}
We note that this mechanism has previously been discussed  in the context of AGN outflows by \cite{1995ApJ...451..510S}. These authors also arrived at similar terminal speed for a radiation driven outflow.

We plot this scaling to compare with the observed correlation of $\beta$ and 
$L_{\rm bol}$ in Figure 2. The dash-dotted, solid and dotted line in Figure 2 correspond to $r_{b,0}=0.2,0.5, 1.0$ pc respectively. We find that this simple theoretical model fits the observed correlation pretty well, which indicates that the absorber systems are likely to be radiation(dust) driven outflows. 

One is then tempted to ask as to how these outflows fit in the unification schemes of AGN. We find that the launching radius of the outflows is the dust sublimation radius, which is also the inner radius of the dusty torus. Inside the torus, the UV photons are quickly reprocessed into IR. Although the IR photons can also drive outflows \citep{2011ApJ...741...29D, 2012arXiv1209.0242S}, however the speeds would not be large, as the IR to dust scattering  cross section is more than an order of magnitude smaller than in UV. One possible way to reconcile this is the following.

Let us suppose that the outflows do not plough through the main body of the torus,  but consist of material lifted from the outer surface of the torus. In that case, as the torus material is dilute and highly porous at its periphery,  the UV photons can in principle travel a large distance without being attenuated and impart their momentum to gas and dust mixture lifted
from the outer surface of the torus. More specifically, in the picture presented in \cite{2000ApJ...545...63E}, the region which we are considering should take place between the BAL envelope and the torus. We note that, this scenario not only gives rise to large velocity outflows, but it may also account for the small fraction ($\lesssim 0.1$) of the QSOs which show these absorbers owing to the fact that the region allowed for the outflows (periphery of the torus) occupies a very small fraction of the viewing angle.   
\section{Fractional number of  absorbers}
Next we study the dependence of  fractional number counts of absorbers as a function of QSO  luminosity. 
 We define the fractional number count as below, 
\begin{equation}
\rm Frac.\ number\ count = \frac{Number\ of\ absorbers\ found}{Number\ of\ QSOs\ searched\ in\ a\  bin} \,. \nonumber
\end{equation}
Again, we limit our analysis to the spectral region with $\beta<0.4$. From our
sample, as described in \S 2,  we can easily estimate the ``Number of  absorbers found" 
in a given luminosity 
bin, having $\beta<0.4$.  However to find the corresponding ``Number of QSOs
searched in a bin",
we  also need to count those QSOs in the parent sample of SDSS-DR7 from which the QSOs with \mgtwo absorbers are selected.  We use the parent catalogue from \citet{shen2011ApJS..194...45S} of which the sample used in this work is a subset.
  Therefore, we estimate the 
``Number of QSOs searched in a bin"  by  using non-BAL QSOs from Shen et al (2012) catalogue, which satisfy the redshift criteria $1.185 \le z_{\rm qso} < 2.21$,
to ensure the absence of  any observational biases (see \S 2).

We plot the fractional number count as a function of luminosity ($L_{\rm bol}$) in Figure 3. The values are shown by filled diamonds whose x-coordinate is the centre of each luminosity bin. We also show the overall average of the sample using a horizontal line, whose value is $\sim0.1$. 
\begin{figure}
\includegraphics[scale=0.4]{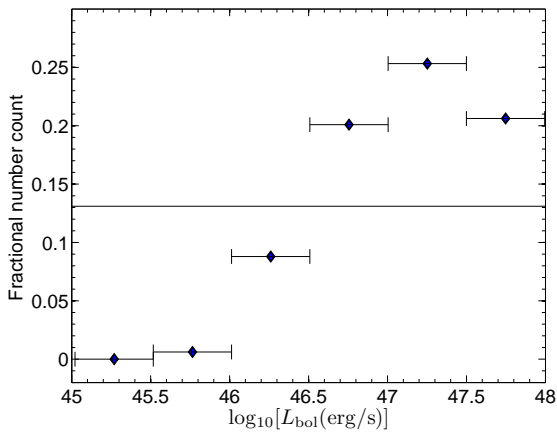}
\caption{The fractional number count as a function of bolometric luminosity is shown using filled diamonds, and the size of the luminosity bin is shown by the horizontal bar. Thin horizontal line represents the average value of fractional number count over the entire sample.}
\end{figure}
We find that the fraction increases steeply  with increasing QSO luminosity and reaches a maximum roughly for $L_{\rm bol} \sim 10^{47}$ erg s$^{-1}$. For $L_{\rm bol}> 10^{47.5}$ erg s$^{-1}$ there is a mild decrease with luminosity, however this decrease is uncertain as in this bin, we have many apparently faint high redshift quasars, for which the signal to noise criterion removes large chunks of spectra and the corresponding absorbers (D. Lawther pvt. comm.). 
We note that the fractional number of absorbers has a contribution from outflowing and intervening systems, which we can not separate here.

\section{Discussion}
We would like to emphasize an important point in connection with our result. There is a general consensus in the literature, which goes along the line that the absorbers which  have $\beta<0.0167$ ($v<5000$ km s$^{-1}$), are associated with the QSO and with $\beta$ higher than this represent the intervening media. 
We stress here that this criterion  is not adequate to denote the associated systems, and the true associated systems can also have $\beta>0.0167$, e.g. the QSO driven high velocity outflows considered here.

To illustrate this, we plot in Figure 4, the ratio of absorbers with $\beta<0.0167$ to the total number of absorbers in a particular luminosity bin as a function of bolometric luminosity.  One can clearly see that lower $\beta$ are possible for only lower luminosity, and vice versa. Firstly, the figure once again confirms that the velocity offset $\beta$ is correlated with luminosity, because low
$\beta$ absorbers appear along the sightlines of low luminosity QSOs. Secondly, this plot, in conjunction with the correlation of $\beta$ with $L_{\rm bol}$, shows that the systems which are  really `{\it associated} ' with the QSOs are spread  all the way from $\beta=0.0$ to $0.4$.

Our results call for a study to separate out the truly associated (outflowing) systems and the intervening ones. Of course, one tedious way to do this is to locate the intervening galaxies in each quasar sightline, however yet another  way can be through the detailed study of line shapes and features arising from outflows and intervening material. We look forward to such a study in the future. 
\begin{figure}
\includegraphics[scale=0.4]{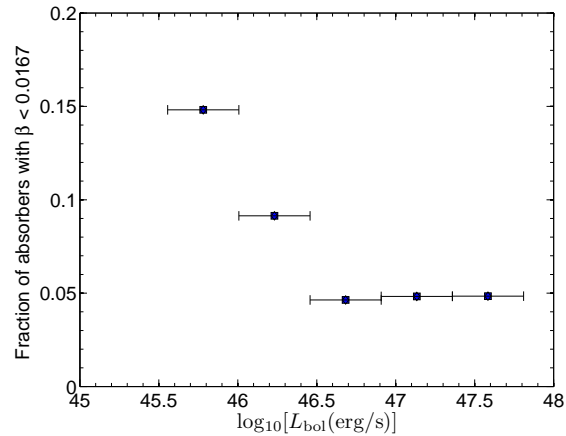}
\caption{Ratio of absorbers with $\beta<0.0167$ to the total number of quasars in a particular luminosity bin is plotted against the bolometric luminosity. The horizontal bar represents the size of the luminosity bin.}
\end{figure}

There is another implication of the observed dependence of fractional number
count of absorbers on QSO luminosity. If one considers a sample of a particular
type of QSOs that covers a restricted luminosity range, then the relative
number of absorbers may differ for different samples, and be different from
the overall average.
If we consider the right side of Figure 3, corresponding to  $L_{\rm bol} > 10^{46.5}$ erg/s, there the fractional number counts  are roughly double of the overall average value of $0.1$. We note
that recent observations of $\sim 45$ blazars \citep{Bergeron2011A&A...525A..51B} report an excess of \mgtwo
absorbers relative to that in QSOs. We speculate here that this excess may also arise from the fact that the blazar sample is small,  and  it may be possible that it is biased towards higher luminosity,  where the fractional number count is larger. It is 
possible  that if the analysis is repeated with a larger sample of blazars then the excess may fade away. In fact, a similar conclusion has been
reached for a sample of FSRQs and $7156$ lobe and core dominated QSOs where in both cases one  finds only a mild excess \citep{Joshi_etal}. In this regard we bring  a recent paper by  \citet{Cucchiara2012arXiv1211.6528C}  to the attention of the reader, regarding the excess seen towards GRBs, where with a large sample of GRBs the puzzle of \mgtwo incidence rate  indeed disappears, and one does not find any excess. 

In summary,  we have found a correlation between the velocity offset of strong \mgtwo absorbers  and the luminosity of QSOs. The velocity offset ($\beta c$) has been found to increase with the luminosity with a power law index $\sim 1/4$. We have found that radiation driven outflows from QSOs can give rise to such a dependence of $\beta$ on $L_{\rm bol}$. These findings lead us to conclude that a significant fraction of strong \mgtwo absorbers (even with $v> 5000$ \kms) along QSO sightlines  may be the AGN driven outflows.

We are grateful to D. Lawther for supplying the redshift path data. We thank an anonymous referee for insightful comments.
\footnotesize{
\bibliography{references}
}

\appendix
\section{The median value of  $\beta$}
From equation (1), we know that $\beta$ mainly depends on the  difference of 
$(1+z_{\rm qso})$ and $(1+z_{\rm abs})$.   
Here we address the question whether or not 
the median value of $\beta$ which we have computed over its 
fixed range varying from $0\hbox{--}0.4$,  depends on the value of $z_{\rm qso}$.
 For maximum value of $\beta=0.4$, the lower limit of $z_{\rm abs}$, using equation (1) is
given by
$
z_{\rm abs}^{\rm min} = (C-1) + C z_{\rm qso} \,,
$
where $C$ is a constant whose value is  0.65 for $\beta = 0.4$.
The median value of absorber redshift ($z_{\rm abs}^{\rm med}$) is the solution of following equation,
\begin{equation}
\int_{z_{\rm abs}^{\rm min}}^{z_{\rm abs}^{\rm med}} \frac{dN}{dz} dz = \frac{1}{2}\int_{z_{\rm abs}^{\rm min}}^{z_{\rm qso}} \frac{dN}{dz} dz
\label{app_med}
\end{equation}
where $dN/dz$ is the number of \mgtwo absorbers per unit redshift. If the absorbers are distributed uniformly and the quantity $dN/dz$ is constant, then from equation (\ref{app_med}) we get,
$
z_{\rm abs}^{\rm med} = ((C-1) + (C+1) z_{\rm qso})/2 \,.
$
Using equation (1) we find the corresponding median value of $\beta$, which is 
$
\beta_{\rm med} \approx 0.19 \,,
$
independent of $z_{\rm qso}$.
However, the observed  $\beta_{\rm med}$ evolves  with luminosity (and $z_{\rm qso}$), which indicates that absorbers are not distributed uniformly.

\label{lastpage} 
\end{document}